\documentclass[10pt,conference]{IEEEtran}
\IEEEoverridecommandlockouts
\usepackage{graphicx}
\usepackage{graphics}
\usepackage{epsfig}
\usepackage{amsfonts}
\usepackage{amssymb}
\usepackage{amscd}
\usepackage{array}
\usepackage{eqparbox}
\usepackage[cmex10]{amsmath}
\newtheorem{theorem}{Theorem}[section]

\newtheorem{note}[theorem]{Note}

\newcommand{\qed}{\nobreak \ifvmode \relax \else
      \ifdim\lastskip<1.5em \hskip-\lastskip
      \hskip1.5em plus0em minus0.5em \fi \nobreak
      \vrule height0.75em width0.5em depth0.25em\fi}
\def\BibTeX{{\rm B\kern-.05em{\sc i\kern-.025em b}\kern-.08em
    T\kern-.1667em\lower.7ex\hbox{E}\kern-.125emX}}

\begin{document}
\title{Secure Broadcasting With Side-Information}
\author{K. G. Nagananda$^1$, Chandra R Murthy$^2$ and Shalinee Kishore$^1$
\\\\
\begin{tabular}[h]{ccc}
  $^1$Dept. of ECE & $^2$Dept. of ECE    \\
  Lehigh University & Indian Institute of Science    \\
  Bethlehem, PA 18105, USA & Bangalore 560012, India \\
  \{kgn209, skishore\}@lehigh.edu & cmurthy@ece.iisc.ernet.in
\end{tabular}
}

\maketitle
\begin{abstract}
In this paper, we derive information-theoretic performance limits for secure and reliable communications over the general two-user discrete memoryless broadcast channel with side-information at the transmitter. The sender wishes to broadcast two independent messages to two receivers, under the constraint that each message should be kept confidential from the unintended receiver. Furthermore, the encoder has side-information - for example, fading in the wireless medium, interference caused by neighboring nodes in the network, etc. - provided to it in a noncausal manner, i.e., before the process of transmission. We derive an inner bound on the capacity region of this channel, by employing an extension of Marton's coding technique used for the classical two-user broadcast channel, in conjunction with a stochastic encoder to satisfy confidentiality constraints. Based on previously known results, we discuss a procedure to present a schematic of the achievable rate region. The rate-penalties for dealing with side-information and confidentiality constraints make the achievable region for this channel strictly smaller than the rate regions of those channels where one or both of these constraints are relaxed.
\end{abstract}

\section{Introduction}\label{sec:introduction}
In the theory of cooperative communications, \emph{side-information} has been used as a basis for user-cooperation, which has been actively pursued as a key enabling technology to meet the demands of higher data-rates and efficient utilization of radio-frequency spectrum. User cooperation is especially popular in wireless networks with multiple nodes, where a particular node expresses its willingness to share its data (or other resources) in a causal or noncausal manner. One such multiple node network is the broadcast channel (BC) \cite{refcoverbroadcast1}, which has received vast attention since its inception into network information theory. Characterization of performance limits for BC has been an active area of research, with Marton deriving the best known inner bound on the capacity region for the general two-user discrete memoryless version of the channel \cite{refmartonbroadcast1}. Some of the most prominent information-theoretic results on BC have been summarized in \cite{refcoverbroadcast2}.

Yet another issue in wireless communications, owing to the broadcast nature of the wireless medium, is related to information security. That is, the broadcast nature of wireless networks facilitates malicious or unauthorized access to confidential data, denial of service attacks, corruption of sensitive data, etc. An information-theoretic approach to address problems related to security has gained rapid momentum, and is commonly referred to as information-theoretic confidentiality or wireless physical-layer security \cite{refliang2}.

\subsection{Our contribution}\label{subsec:ourcontribution}
In this paper, we consider a general two-user BC with $(i)$ side-information at the transmitter and $(ii)$ confidential messages. The sender, denoted $\mathrm{S}$, has two messages $m_1$ and $m_2$ intended for two destinations, denoted $\mathrm{D}_1$ and $\mathrm{D}_2$, respectively, such that $m_1$ (resp. $m_2$) has to be kept confidential from $\mathrm{D}_2$ (resp. $\mathrm{D}_1$). Furthermore, the encoder at $\mathrm{S}$ has noncausal knowledge of random parameters - for example, fading in the wireless medium, interference caused by neighboring nodes in the network, etc. We present an inner bound on the capacity region by deriving a set of achievable rate pairs for secure and reliable communications, by considering the discrete memoryless version of this channel.

The achievability theorem is proved by employing an extension of Marton's coding technique, used to derive a rate region for the general two-user BC, in conjunction with a stochastic encoder at $\mathrm{S}$ to satisfy confidentiality constraints. We also discuss a procedure for presenting a schematic of the achievable rate region; our arguments are motivated by well-known results for Gel'fand-Pinsker's (GP) channel with random parameters \cite{refgelfand} and wiretap channel with side-information \cite{refvinck1}. Results demonstrate that, owing to rate-penalties for dealing with side-information and satisfying confidential constraints, the achievable rate region for our communication setup is strictly smaller than the rate regions of the classical two-user BC and BC with noncausal side-information.

\subsection{Related work}\label{subsec:relatedwork}
An inner bound on the capacity region for BC with noncausal side-information at the transmitter has been presented in \cite{refsteinberg1}, where Marton's achievability scheme has been extended to the case of state-dependent channels. It is also shown that, in the case of Gaussian channels, the capacity region coincides with that of the same channel without states.
In \cite{refsteinberg2}, the degraded BC with random parameters at the encoder are considered under two separate scenarios: When the states are available in a noncausal manner, and when side-information is provided in a causal manner. Capacity bounds are derived for the channel with noncausal states, and the bounds are shown to be tight when the non-degraded user is informed about the channel parameters. For the causal case, a single-letter characterization of the capacity region is derived.

Characterization of performance limits for BC with side-information at the receivers have also been addressed in the literature. For example, in \cite{refkramerbroadcast1}, the capacity region for the general two-user discrete memoryless BC has been derived when each receiver has prior information of the message that it need not decode. This result generalizes to the additive white Gaussian noise channels with average power constraints, and also to the degraded case where one receiver decodes both messages. A slightly different model is considered in \cite{refalon1}, where a sender wishes to broadcast \emph{blocks} of data to multiple receivers, with each receiver having prior side-information consisting of some subset of the other blocks. A bound has been derived on the minimum number of bits to be transmitted in each block, generalizing several coding theoretic parameters related to source, index and network coding. A source coding perspective for BC has been presented in \cite{refsharmabroadcast1}, where rate-distortion functions under fidelity criterion are defined for a BC when side-information of the source is provided at both the encoder and the decoders.

An information-theoretic approach to secure broadcasting was inspired by the pioneering work of Csisz\'{a}r and K\"{o}rner \cite{refcsiszar1}, who derived capacity bounds for the two-user BC, when the sender transmits a private message to $\mathrm{receiver}$ $1$ and a common message to both receivers, while keeping the private message confidential from $\mathrm{receiver}$ $2$. In \cite{refliu1}, capacity bounds have been derived for BC with a sender broadcasting two independent messages to two receivers, by keeping each message confidential from the unintended receiver.

In this paper, we address the problem of information security over a BC with \emph{noncausal side-information at the transmitter}, making this work novel compared to those in the existing literature. The remainder of the paper is organized as follows. In Section \ref{sec:systemmodel}, we introduce the notation used and provide a mathematical model for the discrete memoryless version of the channel considered in this paper. In Section \ref{sec:mainresult}, we describe an inner bound to the capacity region for this channel, present a schematic of the achievable rate region and compare it with some of the results in the existing literature. We conclude the paper in Section \ref{sec:conclusions}. The proof of the achievability theorem is relegated to appendices.

\section{System Model \& Preliminaries}\label{sec:systemmodel}
We denote the two-user broadcast channel with side information and confidential messages by $\mathrm{C}$. Calligraphic letters are used to denote finite sets, with a probability function defined on them. Uppercase letters denote random variables (RV), while boldface uppercase letters denote a sequence of RVs. Lowercase letters are used to denote particular realizations of RVs, and boldface lowercase letters denote $\mathrm{N}-$length vectors. $\mathrm{N}$ is the number of channel uses and $n=1,\dots,\mathrm{N}$ denotes the channel index. Discrete RV $X \in \mathcal{X}$ and $Y_t \in \mathcal{Y}_t$ denote the channel input and outputs, respectively; $t=1,2$ denotes the receiver index. The encoder is supplied with noncausal side-information $\mathbf{w} \in \mathcal{W}^\mathrm{N}$. The channel is assumed to be memoryless and is characterized by the conditional distribution $p(\mathbf{y}_1,\mathbf{y}_2|\mathbf{x},\mathbf{w}) = \prod_{n=1}^{\mathrm{N}}p(y_{1,n},y_{2,n}|x_{n},w_{n})$.

To transmit its messages, $\mathrm{S}$ generates two RVs $M_{t} \in \mathcal{M}_{t}$, where $\mathcal{M}_{t} = \{1,\dots,2^{\mathrm{N}R_{t}}\}$ denotes the set of message indices. Without loss of generality, $2^{\mathrm{N}R_{t}}$ is assumed to be an integer, with $R_{t}$ being the transmission rate intended to $\mathrm{D}_t$. $M_{t}$ denotes the message $\mathrm{S}$ intends to transmit to $\mathrm{D}_t$, and is assumed to be independently generated and uniformly distributed over the finite set $\mathcal{M}_{t}$. Integer $m_{t}$ is a particular realization of $M_{t}$ and denotes the message-index.

For the channel $\mathrm{C}$, a $((2^{\mathrm{N}R_{1}},2^{\mathrm{N}R_{2}}),\mathrm{N},P_e^{(\mathrm{N})})$ code comprises:
\begin{enumerate}
\item  A stochastic encoder, which is defined by the matrix of conditional probabilities $\phi(\mathbf{x}|m_{1},m_{2},\mathbf{w})$, such that $\sum_{\mathbf{x}}\phi(\mathbf{x}|m_{1},m_{2},\mathbf{w}) = 1$. Here, $\phi(\mathbf{x}|m_{1},m_{2},\mathbf{w})$ denotes the probability that a pair of message-indices $(m_{1},m_{2})$ is encoded as $\mathbf{x} \in \mathcal{X}^\mathrm{N}$ to be transmitted by $\mathrm{S}$, in the presence of noncausal side-information $\mathbf{w}$.
\item Two decoders - $g_t: \mathcal{Y}^\mathrm{N}_t \rightarrow \mathcal{M}_{t}$.
\end{enumerate}
The average probability of decoding error for the code, averaged over all codes, is
$P_e^{(\mathrm{N})} = \max \{P_{e,1}^{(\mathrm{N})},P_{e,2}^{(\mathrm{N})}\}$, where,
\begin{eqnarray*}
P_{e,t}^{(\mathrm{N})} = \sum_{\mathbf{m}}\sum_{\mathbf{w}\in \mathcal{W}^\mathrm{N}}\frac{1}{2^{\mathrm{N}[R_{1}+R_{2}]}}\text{Pr}\left[g_t(\mathcal{Y}_t^\mathrm{N}) \neq m_{t}|\mathbf{m},\mathbf{w}~ \text{sent}\right],
\end{eqnarray*}
where $\mathbf{m} = (m_{1},m_{2})$. A rate pair $(R_{1},R_{2})$ is said to be achievable for the channel $\mathrm{C}$, if there exists a sequence of $((2^{\mathrm{N}R_{1}},2^{\mathrm{N}R_{2}}),\mathrm{N},P_e^{(\mathrm{N})})$ codes $\forall \epsilon > 0$ and sufficiently small, such that $P_e^{(\mathrm{N})} \leq \epsilon$ as $\mathrm{N} \rightarrow \infty$ and the following weak-secrecy constraints \cite{refmaurer2} are satisfied:
\begin{eqnarray}
\mathrm{N}R_{1} - H(M_{1}|Y_2) \leq \mathrm{N}\epsilon, \label{eq:security1}   \\
\mathrm{N}R_{2} - H(M_{2}|Y_1) \leq \mathrm{N}\epsilon, \label{eq:security2}
\end{eqnarray}
where $H(\mathrm{\alpha}|\mathrm{\beta})$ is the conditional entropy of $\mathrm{\alpha}$ given $\mathrm{\beta}$. The weak-secrecy rate can be replaced by the \emph{strong-secrecy key rate} without any penalty \cite{refmaurer2}. The capacity region is defined as the closure of the set of all achievable rate pairs $(R_{1},R_{2})$.

\section{Main Result \& Discussion}\label{sec:mainresult}
\begin{figure*}[ht!]
\centering
\begin{eqnarray}
R_1 &\leq& I(V_1;Y_1|U) - \max [I(V_1;Y_2|U,V_2),I(W;V_1|U)],\label{eq:rateregionR1}\\
R_2 &\leq& I(V_2;Y_2|U) - \max [ I(V_2;Y_1|U,V_1),I(W;V_2|U)],\label{eq:rateregionR2}\\
R_1 + R_2 &\leq& I(V_1;Y_1|U) + I(V_2;Y_2|U) - I(V_1;Y_2|U,V_2) - I(V_2;Y_1|U,V_1) - I(V_1;V_2|U) - I(V_1,V_2;W|U). \label{eq:rateregionR1plusR2}
\end{eqnarray}
\hrulefill
\end{figure*}
In this section, we present an achievable rate region the channel $\mathrm{C}$. We also discuss, based on previously known results, the procedure adopted to obtain a schematic of this achievable rate region.

\subsection{An achievable rate region}\label{subsec:rateregion}
Consider the following auxiliary RVs defined on finite sets: $U \in \mathcal{U}$ and $V_t \in \mathcal{V}_t;t=1,2$. Let $\mathcal{P}$ denote the set of all joint probability distributions $p(w,u,v_1,v_2,x,y_1,y_2)$ that is constrained to factor as follows:
\begin{eqnarray*}
p(w,u,v_1,v_2,x,y_1,y_2) = p(w)p(u)p(v_1,v_2|w,u)\\ \times p(x|w,v_1,v_2)p(y_1,y_2|x).
\end{eqnarray*}
For a given $p(.)\in \mathcal{P}$, an achievable rate region for $\mathrm{C}$ is described by the set $\mathfrak{R}_{\text{in}}(p)$, which is defined as the convex-hull of the set of all rate pairs $(R_{1},R_{2})$ that simultaneously satisfy $(\ref{eq:rateregionR1})$ - $(\ref{eq:rateregionR1plusR2})$.
\begin{theorem}\label{thm:achievethmC}
Let $\mathfrak{C}$ denote the capacity region of the channel $\mathrm{C}$.  Let $\mathfrak{R}_{\text{in}} = \bigcup_{p(.)\in \mathcal{P}}\mathfrak{R}_{\text{in}}(p)$. The region $\mathfrak{R}_{\text{in}}$ is an achievable rate region for $\mathrm{C}$, i.e., $\mathfrak{R}_{\text{in}}\subseteq \mathfrak{C}$.
\end{theorem}
The proof of Theorem \ref{thm:achievethmC} can be found in Appendices \ref{appendix:errordecode} and \ref{appendix:errorencode}.

\subsection{Discussion}\label{subsec:discussion}
\begin{figure}[h]
\flushright
\includegraphics[height=2in,width=3.1in]{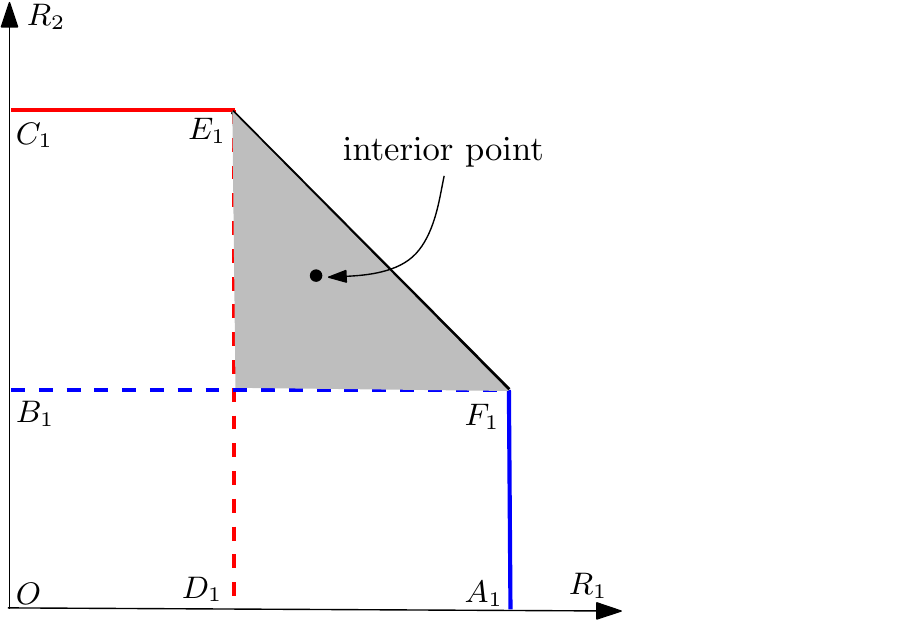}
\caption{Schematic of the rate region for secure BC with side-information}\label{fig:securebcsideinfo_rateregion}
\end{figure}
For the channel $\mathrm{C}$, rate inequalities $(\ref{eq:appendArates})$, constraints $(\ref{eq:appendAR1'})$ - $(\ref{eq:appendAR2*})$ and bounds on the binning rates $(\ref{eq:appendBR1*})$ - $(\ref{eq:appendBR1R2*})$ are combined to obtain the rate region described by $(\ref{eq:rateregionR1})$ - $(\ref{eq:rateregionR1plusR2})$. We employ
now arguments from Gelfand-Pinsker's channel with random parameters \cite{refgelfand} and wiretap channels with side-information \cite{refvinck1} to present a schematic of the rate region (see Fig. \ref{fig:securebcsideinfo_rateregion}).

When $R_2=0$, the channel resembles a wiretap channel with side-information and $\mathrm{S}$ can transmit at the maximum achievable $R_1$ given by $(\ref{eq:rateregionR1})$, denoted by point $A_1$. When $\mathrm{S}$ is transmitting at point $A_1$, the maximum achievable $R_2$ is given by the point $B_1 \equiv I(V_2;Y_2|U) - I(V_2;Y_1|U,V_1) - \max [I(V_1;V_2|U) ,I(W;V_2|U)]$; this is obtained by treating the channel as a wiretap channel with side-information. Therefore, the rectangle $OA_1F_1B_1$ is achievable. By flipping $R_1$ and $R_2$ and following similar arguments, the points $C_1$, given by $(\ref{eq:rateregionR2})$, and $D_1 \equiv I(V_1;Y_1|U) - I(V_1;Y_2|U,V_2) - \max [I(V_1;V_2|U) ,I(W;V_1|U)]$ are achievable. Hence, the rectangle $OC_1E_1D_1$ is also achievable. Since the points $E_1$ and $F_1$ are shown to be achievable, any point which lies on the line $E_1F_1$ can also be achieved by deriving a bound on the binning rates (see $(\ref{eq:appendBR1R2*})$, Appendix \ref{appendix:errorencode}). This leads to a sum rate bound given by $(\ref{eq:rateregionR1plusR2})$. Finally, owing to convexity of the rate region, any point in the interior of the line $E_1F_1$ is also achievable. Therefore, an achievable rate region for $\mathrm{C}$ is described by the pentagon $OA_1F_1E_1C_1$.
\begin{figure}[t]
\flushright
\includegraphics[height=2in,width=3.2in]{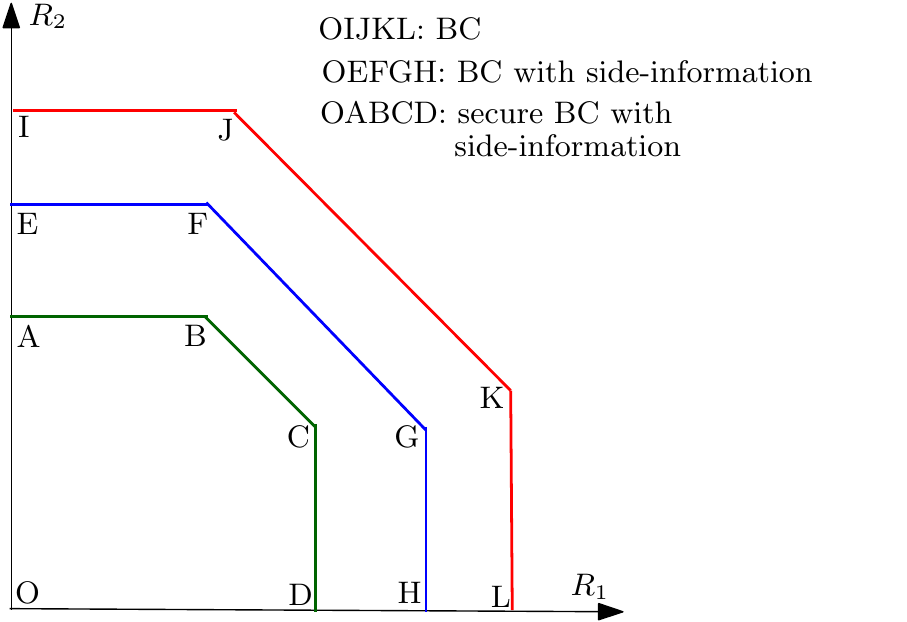}
\caption{Schematic of rate regions for $(i)$ BC, $(ii)$ BC with side-information and $(iii)$ secure BC with side-information. The regions are not to scale and are depicted here for illustrative purposes only.}\label{fig:bc_compare}
\end{figure}

If the confidentiality constraints $(\ref{eq:security1})$ - $(\ref{eq:security2})$ are relaxed, the channel $\mathrm{C}$ reduces to a broadcast channel with side-information whose rate region is the pentagon OEFGH (see Fig. \ref{fig:bc_compare}), first characterized by Steinberg and Shamai \cite{refsteinberg1}. It is described by the convex-hull of the set of all rate pairs $(R_1,R_2)$ that satisfy the following inequalities:
\begin{eqnarray}
R_1 &\leq& I(V_1;Y_1) - I(W;V_1),\label{eq:rateregionBCsideinfoR1}\\
R_2 &\leq& I(V_2;Y_2) - I(W;V_2),\label{eq:rateregionBCsideinfoR2}\\
\nonumber R_1 + R_2 &\leq& I(V_1;Y_1) + I(V_2;Y_2)\\ && - I(V_1;V_2) - I(V_1,V_2;W). \label{eq:rateregionBCsideinfoR1plusR2}
\end{eqnarray}
Further, in the absence of side-information, i.e., $\mathcal{W}=\{\phi\}$, the channel reduces to the classical two-user broadcast channel whose rate region is the pentagon OIJKL, first characterized by Marton \cite{refmartonbroadcast1}. It is described by the convex-hull of the set of all rate pairs $(R_1,R_2)$ that satisfy the following inequalities:
\begin{eqnarray}
R_1 &\leq& I(V_1;Y_1),\label{eq:rateregionBCR1}\\
R_2 &\leq& I(V_2;Y_2),\label{eq:rateregionBCR2}\\
R_1 + R_2 &\leq& I(V_1;Y_1) + I(V_2;Y_2) - I(V_1;V_2). \label{eq:rateregionBCR1plusR2}
\end{eqnarray}
Lastly, if the encoder satisfies confidentiality constraints in the absence of side-information, the channel $\mathrm{C}$ reduces to a broadcast channel with confidential messages whose rate region was first characterized by Liu et. all \cite{refliu1}. It is described by the convex-hull of the set of all rate pairs $(R_1,R_2)$ that satisfy the following inequalities:
\begin{eqnarray}
R_1 \leq I(V_1;Y_1|U) - I(V_1;Y_2|U) - I(V_1;V_2|U),\label{eq:rateregionBCconfR1}\\
R_2 \leq I(V_2;Y_2|U) - I(V_2;Y_1|U) - I(V_1;V_2|U).\label{eq:rateregionBCconfR2}
\end{eqnarray}
\begin{note}
The rate region for BC with side-information $(\ref{eq:rateregionBCsideinfoR1})$ - $(\ref{eq:rateregionBCsideinfoR1plusR2})$ is smaller than that of the classical BC $(\ref{eq:rateregionBCR1})$ - $(\ref{eq:rateregionBCR1plusR2})$ , due to the rate-penalty for side-information. For $I(V_1;Y_2|U,V_2)>I(W;V_1|U)$ and $I(V_2;Y_1|U,V_1)>I(W;V_2|U)$, the achievable region of channel $\mathrm{C}$ is smaller than that for BC with side-information. The rate region for $\mathrm{C}$ is at most as large as that for BC with side-information. This provides the necessary intuition for the dimensions (though, they are not to-scale in Fig. \ref{fig:bc_compare}) of the pentagon OIJKL, which subsumes OEFGH which further subsumes OABCD.
\end{note}
\begin{note}
All the above described rate regions can also be obtained by arguing along lines similar to those used to obtain the rate region for $\mathrm{C}$, described by the pentagon OABCD $\equiv OA_1F_1E_1C_1$.
\end{note}
\begin{note}
Outer bounds for the channel model in this paper, as well as for the model considered in \cite{refsteinberg1}, have been derived in a followup paper \cite{refnandaicc2012BC2}.
\end{note}

\section{Conclusions}\label{sec:conclusions}
We presented an inner bound on the capacity region of a two-user BC with ($i)$ noncausal side-information at the encoder and $(ii)$ confidentiality constraints, such that each message is kept secret from the unintended receiver. The achievability proof involved extension of techniques from Marton's coding scheme for the general BC and a stochastic encoder to achieve information secrecy. We also discussed a simple procedure to present a schematic of the rate region for this channel and argued that, due to rate-penalties for using side-information and maintaining information-secrecy, the achievable region is strictly smaller than the rate regions for channels where these constraints are relaxed.


\appendices
\section{}\label{appendix:errordecode}
Here, we present the codebook construction, upper bound the probability of decoding errors and perform equivocation calculations to show that the code satisfies confidentiality constraints. We denote by $A^{(\mathrm{N})}_{\epsilon}(P_{\mathrm{x}})$ an $\epsilon$-typical set comprising $\mathrm{N}-$sequences picked from a distribution $P_{\mathrm{x}}$. The encoder at $\mathrm{S}$ is given, in a noncausal manner, the $\mathrm{N}-$sequence $\mathbf{w}$ picked from a distribution $P(\mathbf{w})=\prod_{n=1}^{\mathrm{N}}P(w_n)$. Generate a typical $\mathrm{N}-$sequence $\mathbf{u}$, known to all nodes in the network, picked from the distribution $P(\mathbf{u})=\prod_{n=1}^{\mathrm{N}}P(u_n)$. Generate $2^{\mathrm{N}[R_t+R^{'}_t+R_t^{\ast}]}$ independent $\mathrm{N}-$sequences $\mathbf{v}_t(i_t,j_t,k_t)$, picked from the distribution $P(\mathbf{v}_t|\mathbf{u})=\prod_{n=1}^{\mathrm{N}}P(v_{t,n}|u_n)$. Here, $i_t\in\{1,\dots,2^{\mathrm{N}R_t}\}$; $j_t\in\{1,\dots,2^{\mathrm{N}R^{'}_t}\}$; $k_t\in\{1,\dots,2^{\mathrm{N}R_t^{\ast}}\}$. Without loss of generality, $2^{\mathrm{N}R_t}$, $2^{\mathrm{N}R^{'}_t}$ and $2^{\mathrm{N}R_t^{\ast}}$ are considered to be integers. The following double-binning scheme is employed:
\begin{enumerate}
\item Uniformly distribute $2^{\mathrm{N}[R_t+R^{'}_t+R_t^{\ast}]}$ $\mathrm{N}-$sequences into $2^{\mathrm{N}R_t}$ bins, so that each bin indexed by $i_t$ comprises $2^{\mathrm{N}[R^{'}_t+R_t^{\ast}]}$ $\mathrm{N}-$sequences.
\item Uniformly distribute $2^{\mathrm{N}[R^{'}_t+R_t^{\ast}]}$ $\mathrm{N}-$sequences into $2^{\mathrm{N}R^{'}_t}$ sub-bins indexed by $(i_t,j_t)$, so that each bin comprises $2^{\mathrm{N}R_t^{\ast}}$ $\mathrm{N}-$sequences.
\end{enumerate}

To send the message pair $(m_1,m_2)$, $\mathrm{S}$ employs a stochastic encoder. In the bin indexed by $i_t$, \emph{randomly} pick a sub-bin indexed $(i_t,j_t)$. The encoder then looks for a pair $(k_1,k_2)$ that satisfies the following joint typicality condition $E_{\mathrm{S}} \triangleq $
\begin{eqnarray}
(\mathbf{w},\mathbf{v}_1(i_1,j_1,k_1),\mathbf{v}_2(i_2,j_2,k_2))\in A^{(N)}_{\epsilon}(P_{W,V_1,V_2|U}). \label{eq:appendencode}
\end{eqnarray}
The channel input is an $\mathrm{N}-$sequence $\mathbf{x}$ picked from the distribution $P(\mathbf{x}|\mathbf{w},\mathbf{v}_1,\mathbf{v}_2)=\prod_{n=1}^{\mathrm{N}}P(x_n|w_n,v_{1,n},v_{2,n})$.

At the destination $\mathrm{D}_t$, given $\mathbf{u}$, the decoder picks $k_t$ that satisfies the following joint typicality condition:
\begin{eqnarray}
(\mathbf{v}_t(i_t,j_t,k_t),\mathbf{y}_t)\in A^{(N)}_{\epsilon}(P_{V_t,Y_t|U}). \label{eq:appenddecode}
\end{eqnarray}

An error is declared at decoder of $\mathrm{D}_t$ if it not possible to find an integer $\hat{i}_t$ to satisfy the condition $E_{\mathrm{D}_t} \triangleq \{\mathbf{v}_t(\hat{i}_t,j_t,k_t),\mathbf{y}_t)\in A^{(\mathrm{N})}_{\epsilon}(P_{V_t,Y_t|U})\}$. From union of events bound, the probability of decoder error at $\mathrm{D}_t$ can be upper bounded as follows:
\begin{eqnarray*}
P^{(\mathrm{N})}_{e,\mathrm{D}_t} \leq P(E^c_{\mathrm{D}_t}|E_{\mathrm{S}}) + \sum_{\hat{i}_t\neq i_t}\sum_{j_t,k_t}P(E_{\mathrm{D}_t}|E_{\mathrm{S}}).
\end{eqnarray*}
From the asymptotic equipartition property (AEP) \cite{refcoverbook}, $\forall \epsilon > 0$ and sufficiently small; and for large $\mathrm{N}$, $P(E^c_{\mathrm{D}_t}|E_{\mathrm{S}}) \leq \epsilon$ and for $\hat{i}_t\neq i_t$
\begin{eqnarray*}
P(E_{\mathrm{D}_t}|E_{\mathrm{S}}) \leq 2^{-\mathrm{N}[I(V_t;Y_t|U)-\epsilon]}.
\end{eqnarray*}
Therefore, we have
\begin{eqnarray*}
P^{(\mathrm{N})}_{e,\mathrm{D}_t} \leq \epsilon + 2^{\mathrm{N}[R_t+R^{'}_t+R_t^{\ast}]}2^{-\mathrm{N}[I(V_t;Y_t|U)-\epsilon]}.
\end{eqnarray*}
For any $\epsilon_0 >0$ and sufficiently small for large $\mathrm{N}$ , $P^{(\mathrm{N})}_{e,\mathrm{D}_t} \leq \epsilon_0$ if
\begin{eqnarray}
R_t+R^{'}_t+R_t^{\ast} < I(V_t;Y_t|U). \label{eq:appendArates}
\end{eqnarray}
The equivocation at the decoder of $\mathrm{D}_2$ is calculated by first considering the following lower bound:
\begin{eqnarray}
H(M_1|\mathbf{Y}_2) \geq H(M_1|\mathbf{Y}_2,\mathbf{U},\mathbf{V}_2).\label{eq:appendAequiv1}
\end{eqnarray}
Following the procedure in \cite[Section V-B]{refliu1} and using the fact that $M_1 \rightarrow (\mathbf{U},\mathbf{V}_1,\mathbf{V}_2) \rightarrow \mathbf{Y}_2$ forms a Markov chain, $(\ref{eq:appendAequiv1})$ becomes
\begin{eqnarray}
\nonumber H(M_1|\mathbf{Y}_2) \geq H(\mathbf{V}_1|\mathbf{U}) - I(\mathbf{V}_1;\mathbf{V}_2|\mathbf{U})\\
- H(\mathbf{V}_1|M_1,\mathbf{U},\mathbf{V}_2,\mathbf{Y}_2) -
I(\mathbf{V}_1;\mathbf{Y}_2|\mathbf{U},\mathbf{V}_2). \label{eq:appendAequiv2}
\end{eqnarray}
Let us consider $\epsilon_l; l=1,\dots,10$, s.t. $\epsilon_l > 0$ and sufficiently small for large $\mathrm{N}$. Let us consider now each term in $(\ref{eq:appendAequiv2})$:
\begin{enumerate}
\item $H(\mathbf{V}_1|\mathbf{U}) \stackrel{(a)}{=} \mathrm{N}[R_1+R'_1+R^{\ast}_1]$,
\item $I(\mathbf{V}_1;\mathbf{V}_2|\mathbf{U})\stackrel{(b)}{=}\mathrm{N}I(V_1;V_2|U)+\mathrm{N}\epsilon_1$,
\item $H(\mathbf{V}_1|M_1,\mathbf{U},\mathbf{V}_2,\mathbf{Y}_2) \stackrel{(c)}{\leq} \mathrm{N}\epsilon_2$,
\item $I(\mathbf{V}_1;\mathbf{Y}_2|\mathbf{U},\mathbf{V}_2) \stackrel{(d)}{=}\mathrm{N}I(V_1;Y_2|U,V_2)+\mathrm{N}\epsilon_3$,
\end{enumerate}
where $(a)$ follows from the codebook construction, $(b)$ and $(d)$ follow from standard techniques and $(c)$ is proved in Appendix \ref{appendix:lemmaproof}. We follow a similar procedure to calculate the equivocation at the decoder at $\mathrm{D}_1$. Finally, the security constraints $(\ref{eq:security1})$ and $(\ref{eq:security2})$ are satisfied by letting
\begin{eqnarray}
R'_1 &=& I(V_1;Y_2|U,V_2) - \epsilon_4,\label{eq:appendAR1'}\\
R_1^{\ast} &=& I(V_1;V_2|U) - \epsilon_5,\label{eq:appendAR1*}\\
R'_2 &=& I(V_2;Y_1|U,V_1) - \epsilon_6,\label{eq:appendAR2'}\\
R_2^{\ast} &=& I(V_1;V_2|U) - \epsilon_7. \label{eq:appendAR2*}
\end{eqnarray}

\section{}\label{appendix:errorencode}
Here, we upper bound the probability of encoder error. An error is declared at the encoder of $\mathrm{S}$ if it is not possible to find a pair $(k_1,k_2)$ to satisfy the condition $E_{\mathrm{S}} \triangleq \{(\mathbf{w},\mathbf{v}_1(i_1,j_1,k_1),\mathbf{v}_2(i_2,j_2,k_2))\in A^{(\mathrm{N})}_{\epsilon}(P_{W,V_1,V_2|U}\}$. Let $P_{e,E_{\mathrm{S}}}$ denote the probability of error at the encoder, i.e., $P_{e,E_{\mathrm{S}}} \triangleq \text{Pr}(E_{\mathrm{S}}^c)$. Let $\mathrm{I}$ be an indicator RV that the event $E_{\mathrm{S}}$ has occurred. Let $Q = \sum_{k_1,k_2} \mathrm{I}$, $\bar{Q} = \mathbb{E}[Q]$ and $\text{Var}[Q] = \mathbb{E}[(Q-\bar{Q})^2]$, where $\mathbb{E}(.)$ denotes the expectation operator. $P_{e,E_{\mathrm{S}}}$ can be upper bounded as follows:
\begin{eqnarray}
P_{e,E_{\mathrm{S}}} = \text{Pr}(Q=0) \stackrel{(e)}{\leq} \text{Var}[Q]/\bar{Q}^2, \label{eq:appendPeS}
\end{eqnarray}
where $(e)$ follows from Markov's inequality for non-negative RVs. Consider now
\begin{eqnarray*}
\bar{Q} &=& \sum_{k_1,k_2}\mathbb{E}(\mathrm{I}) = \sum_{k_1,k_2}\text{Pr}(E_{\mathrm{S}})\\
&\geq& \sum_{k_1,k_2}(1-\delta^{(\mathrm{N})})2^{-\mathrm{N}[I(V_1;V_2|U)+I(V_1,V_2;W|U)+4\epsilon]}\\
&=& (1-\delta^{(\mathrm{N})})2^{-\mathrm{N}[R^{\ast}_1+R^{\ast}_2 - I(V_1;V_2|U) - I(V_1,V_2;W|U) - 4\epsilon]}.
\end{eqnarray*}
Next, consider
\begin{eqnarray*}
\text{Var}[Q] = \sum_{k_1,k_2}\sum_{k'_1,k'_2}\{\mathbb{E}[\mathrm{I}(k_1,k_2)\mathrm{I}(k'_1,k'_2)]\\ - \mathbb{E}[\mathrm{I}(k_1,k_2)]\mathbb{E}\mathrm{I}(k'_1,k'_2)] \}.
\end{eqnarray*}
We have the following four cases:
\begin{enumerate}
\item If $k'_1\neq k_1$ and $k'_2\neq k_2$, then $\mathrm{I}(k_1,k_2)$ and $\mathrm{I}(k'_1,k'_2)$ are independent and $\text{Var}[Q]=0$.
\item If $k'_1 = k_1$ and $k'_2 = k_2$, then $\mathbb{E}[\mathrm{I}(k_1,k_2)\mathrm{I}(k'_1,k'_2)] = \mathbb{E}[\mathrm{I}(k_1,k_2)]\leq 2^{-\mathrm{N}[I(V_1;V_2|U)+I(V_1,V_2;W|U)-4\epsilon]}$.
\item If $k'_1 \neq k_1$ and $k'_2 = k_2$, then $\mathbb{E}[\mathrm{I}(k_1,k_2)\mathrm{I}(k'_1,k'_2)] \leq 2^{-\mathrm{N}[I(V_1;V_2|U)+I(V_1,V_2;W|U)+I(V_1;V_2,W|U)-6\epsilon]}$.
\item If $k'_1 = k_1$ and $k'_2 \neq k_2$, then $\mathbb{E}[\mathrm{I}(k_1,k_2)\mathrm{I}(k'_1,k'_2)] \leq 2^{-\mathrm{N}[I(V_1;V_2|U)+I(V_1,V_2;W|U)+I(V_2;V_1,W|U)-6\epsilon]}$.
\end{enumerate}

Substituting for $\bar{Q}$ and $\text{Var}[Q]$ in $(\ref{eq:appendPeS})$, we can show that $P(E_{\mathrm{S}}) \leq \delta^{(\mathrm{N})}$, $\forall \delta^{(\mathrm{N})} > 0$ and sufficiently small; and for $\mathrm{N}$ large, if the following conditions are simultaneously satisfied:
\begin{eqnarray}
R_1^{\ast} &>& I(W;V_1|U) - \epsilon_8,\label{eq:appendBR1*}\\
R_2^{\ast} &>& I(W;V_2|U) - \epsilon_9,\label{eq:appendBR2*}\\
R_1^{\ast} + R_2^{\ast} &>& I(V_1;V_2|U) + I(V_1,V_2;W|U) - \epsilon_{10}. \label{eq:appendBR1R2*}
\end{eqnarray}

\section{}\label{appendix:lemmaproof}
Here, we prove that $H(\mathbf{V}_1|M_1,\mathbf{U},\mathbf{V}_2,\mathbf{Y}_2) \leq \mathrm{N}\epsilon_2$. This proof is lifted from \cite[Lemma 2]{refliu1} and is provided here for the sake of completeness. Given $M_1=m_1$, the decoder at $\mathrm{D}_2$ chooses $j_1$ and any $k_1$ such that the following typicality condition is satisfied: $\tilde{E} = \{(\mathbf{v}_1(i_1,j_1,k_1),\mathbf{y}_2) \in A^{(\mathrm{N})}_{\epsilon}(P_{V_1,Y_2|U,V_2})\}$. Note that, since $\mathrm{S}$ employs a stochastic encoder, the decoder of $\mathrm{D}_2$ is uncertain about the sub-bin index $j_1$. Let $P^{(\mathrm{N})}_{e,2}$ denote the average probability of error of decoding $j_1$ at $\mathrm{D}_2$. Therefore, we have
\begin{eqnarray*}
P^{(\mathrm{N})}_{e,2} \leq P(\tilde{E}^c|m_{1}~\text{sent}) + \sum_{j_1}P(\tilde{E}|m_{1}~\text{sent}),
\end{eqnarray*}
where $\tilde{E}^c \triangleq \{(\mathbf{v}_1,\mathbf{y}_2) \notin A^{(\mathrm{N})}_{\epsilon}(P_{V_1,Y_2|U,V_2})\}$. From joint AEP \cite{refcoverbook}, $P(\tilde{E}^c|m_{1}~\text{sent})\leq \epsilon$, $\forall \epsilon > 0$ and sufficiently small for large $\mathrm{N}$. And,
\begin{eqnarray*}
P(\tilde{E}|m_{1}~\text{sent}) \leq 2^{-\mathrm{N}[I(V_1;Y_2|U,V_2)-\epsilon]}.
\end{eqnarray*}
Therefore,
\begin{eqnarray*}
P^{(\mathrm{N})}_{e,2} \leq \epsilon + 2^{\mathrm{N}R'_{1}}2^{-\mathrm{N}[I(V_1;Y_2|U,V_2)-\epsilon]}.
\end{eqnarray*}
From equivocation calculations, $R'_{1} = I(V_1;Y_2|U,V_2)-\epsilon_4$. Picking $\epsilon_4 > \epsilon$, we get $P^{(\mathrm{N})}_{e,2} \leq \epsilon$. Next, from Fano's inequality \cite{refcoverbook}, we have
\begin{eqnarray*}
\frac{1}{\mathrm{N}}H(\mathbf{V}_1|M_{1}=m_{1},\mathbf{U},\mathbf{V}_2,\mathbf{Y}_2)\leq \frac{1}{\mathrm{N}}\left[1 + P^{(\mathrm{N})}_{e,2}R'_{1}\right]\\
\leq \frac{1}{\mathrm{N}} + \epsilon I(V_1;Y_2|U,V_2)
\triangleq \epsilon_2.
\end{eqnarray*}
Finally,
\begin{eqnarray*}
\frac{1}{\mathrm{N}}H(\mathbf{V}_1|M_{1},\mathbf{U},\mathbf{V}_2,\mathbf{Y}_2) \leq \\ \frac{1}{\mathrm{N}}\sum_{m_{1}}P(M_{1}=m_{1})H(\mathbf{V}_1|M_{1}=m_{1},\mathbf{U},\mathbf{V}_2,\mathbf{Y}_2)
\leq \epsilon_2.
\end{eqnarray*}

\bibliographystyle{IEEEtran}
\bibliography{IEEEabrv,icc2012}

\begin{thebibliography}{10}
\providecommand{\url}[1]{#1}
\csname url@samestyle\endcsname
\providecommand{\newblock}{\relax}
\providecommand{\bibinfo}[2]{#2}
\providecommand{\BIBentrySTDinterwordspacing}{\spaceskip=0pt\relax}
\providecommand{\BIBentryALTinterwordstretchfactor}{4}
\providecommand{\BIBentryALTinterwordspacing}{\spaceskip=\fontdimen2\font plus
\BIBentryALTinterwordstretchfactor\fontdimen3\font minus
  \fontdimen4\font\relax}
\providecommand{\BIBforeignlanguage}[2]{{%
\expandafter\ifx\csname l@#1\endcsname\relax
\typeout{** WARNING: IEEEtran.bst: No hyphenation pattern has been}%
\typeout{** loaded for the language `#1'. Using the pattern for}%
\typeout{** the default language instead.}%
\else
\language=\csname l@#1\endcsname
\fi
#2}}
\providecommand{\BIBdecl}{\relax}
\BIBdecl

\bibitem{refcoverbroadcast1}
T.~Cover, ``Broadcast channels,'' \emph{IEEE Trans. Inf. Theory}, vol.~18,
  no.~1, pp. 2--14, Jan. 1972.

\bibitem{refmartonbroadcast1}
K.~Marton, ``A coding theorem for the discrete memoryless broadcast channel,''
  \emph{IEEE Trans. Inf. Theory}, vol.~25, no.~3, pp. 306--311, May 1979.

\bibitem{refcoverbroadcast2}
T.~Cover, ``Comments on broadcast channels,'' \emph{IEEE Trans. Inf. Theory},
  vol.~44, no.~6, pp. 2524--2530, Oct. 1998.

\bibitem{refliang2}
Y.~Liang, H.~V. Poor, and S.~Shamai~(Shitz), ``Information theoretic
  security,'' \emph{Found. Trends Commun. Inf. Theory}, vol.~5, no.~4, pp.
  355--580, Apr. 2009.

\bibitem{refgelfand}
S.~Gel'fand and M.~Pinsker, ``Coding for channels with random parameters,''
  \emph{Probl. Contr. and Inf. Theory}, vol.~9, no.~1, pp. 19--31, 1980.

\bibitem{refvinck1}
Y.~Chen and A.~J. Han~Vinck, ``Wiretap channel with side information,''
  \emph{IEEE Trans. Inf. Theory}, vol.~54, no.~1, pp. 395--402, Jan. 2008.

\bibitem{refsteinberg1}
Y.~Steinberg and S.~Shamai~(Shitz), ``Achievable rates for the broadcast
  channel with states known at the transmitter,'' in \emph{Proc. IEEE Int.
  Symp. Inf. Theory}, Adelaide, SA, Sep. 2005, pp. 2184--2188.

\bibitem{refsteinberg2}
Y.~Steinberg, ``Coding for the degraded broadcast channel with random
  parameters, with causal and noncausal side information,'' \emph{IEEE Trans},
  vol.~51, no.~8, pp. 2867--2877, Aug. 2005.

\bibitem{refkramerbroadcast1}
G.~Kramer and S.~Shamai~(Shitz), ``Capacity for classes of broadcast channels
  with receiver side information,'' in \emph{Proc. IEEE Inf. Theory Workshop},
  Tahoe City, CA, Sep. 2007, pp. 313--318.

\bibitem{refalon1}
N.~Alon, E.~Lubetzky, U.~Stav, A.~Weinstein, and A.~Hassidim, ``Broadcasting
  with side information,'' in \emph{Proc. IEEE $49^{th}$ Annual Symp.
  Foundations Comp. Sci.}, Philadelphia, PA, Oct. 2008, pp. 823--832.

\bibitem{refsharmabroadcast1}
B.~D. Sharma and V.~Priya, ``On broadcast channels with side information under
  fidelity criteria,'' \emph{Kybernetika}, vol.~19, no.~1, pp. 27--41, 1983.

\bibitem{refcsiszar1}
I.~Csisz\'{a}r and J.~K\"{o}rner, ``Broadcast channels with confidential
  messages,'' \emph{IEEE Trans. Inf. Theory}, vol. \mbox{IT}-24, no.~3, pp.
  339--348, May 1978.

\bibitem{refliu1}
R.~Liu, I.~Mari\'{c}, P.~Spasojevi\'{c}, and R.~D.Yates, ``Discrete memoryless
  interference and broadcast channels with confidential messages:
  \mbox{Secrecy} rate regions,'' \emph{IEEE Trans. Inf. Theory}, vol.~54,
  no.~6, pp. 2493--2507, Jun. 2008.

\bibitem{refmaurer2}
U.~Maurer and S.~Wolf, ``Information-theoretic key agreement: From weak to
  strong secrecy for free,'' in \emph{Proc. $19^{\mbox{th}}$ Int. Conf. Theory
  App. Crypt. Tech.}, Bruges, Belgium, 2000, pp. 351--368.

\bibitem{refnandaicc2012BC2}
\BIBentryALTinterwordspacing
K.~G. Nagananda, C.~R. Murthy, and S.~Kishore, ``Two classes of broadcast
  channels with side-information: Capacity outer bounds,'' Sep. 2011, submitted
  to IEEE Int. Conf. Comm. [Online]. Available:
  \url{http://arxiv.org/abs/1109.2782}
\BIBentrySTDinterwordspacing

\bibitem{refcoverbook}
T.~Cover and J.~Thomas, \emph{Elements of Information Theory}, 2nd~ed.\hskip
  1em plus 0.5em minus 0.4em\relax New York: Wiley-Interscience, 2006.

\end{thebibliography}

\end{document}